\documentclass[%
unsortedaddress,
preprint,
 amsmath,amssymb,
 aps,
floatfix,
]{revtex4-1}

\usepackage{pdfpages}
\usepackage{pgffor}

\usepackage{xcolor}
\usepackage{graphicx}%
\usepackage{dcolumn}%
\usepackage{bm}%
\usepackage{comment}
\makeatletter
\AtBeginDocument{\let\LS@rot\@undefined}
\makeatother

\newcommand{\si}{Supplemental Information}

\begin{document}

\preprint{}

\title{
Evidence for supercritical behavior \\of high-pressure liquid hydrogen
}%

\author{Bingqing Cheng}
\email{bc509@cam.ac.uk} 
 \affiliation{Laboratory of Computational Science and Modeling, Institute of Materials, {\'E}cole Polytechnique F{\'e}d{\'e}rale de Lausanne, 1015 Lausanne, Switzerland}%
 \affiliation{Trinity College, the University of Cambridge, UK}%
 
\author{Guglielmo Mazzola}
\affiliation{IBM Research Zurich, S\"aumerstrasse 4, 8803 R\"uschlikon, Switzerland
}%

\author{Michele Ceriotti}
\affiliation{Laboratory of Computational Science and Modeling, Institute of Materials, {\'E}cole Polytechnique F{\'e}d{\'e}rale de Lausanne, 1015 Lausanne, Switzerland}%

\date{\today}%

\begin{abstract}

Hydrogen exhibits unusual behaviors at megabar pressures, with consequences for 
planetary science, condensed matter physics and materials science. 
Experiments at such extreme conditions are challenging, often resulting in hard-to-interpret and controversial observations. 
We present a theoretical study of the phase diagram of dense hydrogen, using machine learning to overcome time and length scale limitations while describing accurately interatomic forces.
We reproduce the re-entrant melting behavior and the polymorphism of the solid phase. 
In simulations based on the machine learning potential
we find evidence for continuous metallization in the liquid, as a first-order liquid-liquid transition is pre-empted by freezing. This suggests a smooth transition between insulating and metallic layers in giant gas planets, and reconciles existing discrepancies between experiments as a manifestation of supercritical behavior.
\end{abstract}

\pacs{Valid PACS appear here}%
\maketitle

Hydrogen, the most abundant and simplest element in the universe,
develops a remarkably complex behavior upon compression ~\cite{mcmahon_properties_2012}.
Almost a century ago, Wigner predicted the dissociation and metallization of solid hydrogen at megabar pressures~\cite{wigner1935}.
Since then, unrelenting effort has been made to rationalize the many unusual properties of dense hydrogen, including a rich and poorly understood solid polymorphism~\cite{mcmahon_properties_2012,howie_mixed_2012,zha_high-pressure_2013,dalladay2016evidence}, anomalous melting line~\cite{Bonev2004}, and the possible transition to a superconducting state~\cite{PhysRevLett.21.1748}.

\begin{figure*}[hbt]
\includegraphics[width=1.0\textwidth]{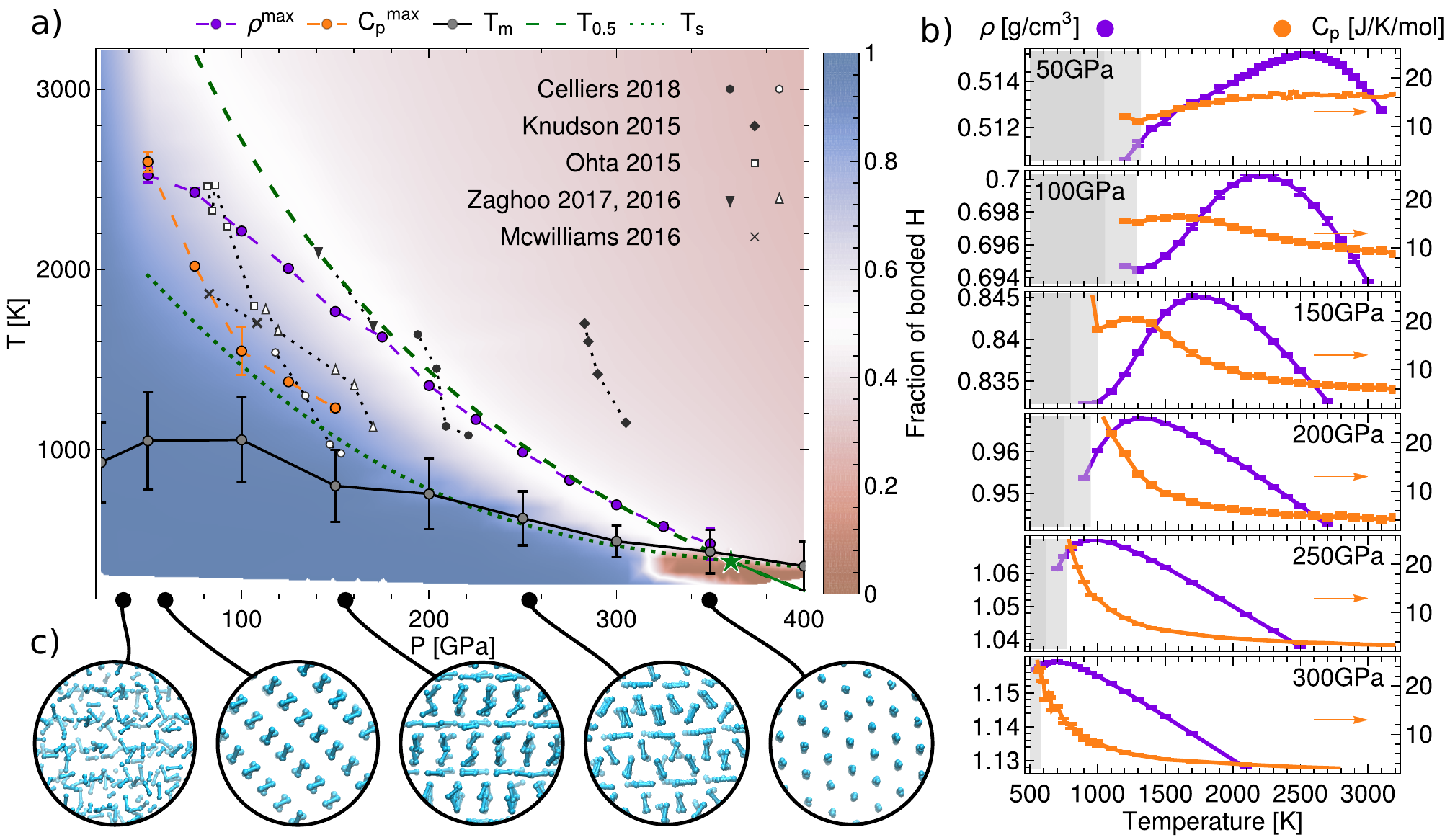}\caption{
Panel (a): High-pressure hydrogen phase diagram.
The color scheme indicates the fraction of molecular hydrogen.
The area below the black line connecting grey dots is the solid hydrogen region, and the upper and the lower bound of the estimated solid-liquid coexistence temperatures are denoted by error bars.
The purple dots indicate the temperatures of maximum density ($\rho$) at constant pressures,
and the orange dots show the locations of molar heat capacity ($c_p$) maxima at different constant pressure conditions.
The dashed and dotted green curve is the solution model prediction of the coexistence line of atomic and molecular fluid (i.e. the fraction of bonded H is 50\%),
and the phase separation temperature at different pressure is plotted using a dashed gray curve.
The intersect between the dashed and the dotted green curves, denoted with a green star, is the predicted location of the critical point of liquid-liquid transition.
The experimental results were taken from Ref.~\citenum{knudson2015direct,Ohta2015,McWilliams2016,Zaghoo2016,Zaghoo2017,Celliers677}.
\\
Panel (b): The purple curves show the density isobar,
and the orange curves show the molar heat capacity ($c_p$) at different pressures.
The shaded regions indicate the conditions under which solid phases are stable, corresponding to the solid-liquid coexistence line shown in panel (a).\\
Panel (c): Crystalline structures of solid hydrogen, obtained at the end of the quenching simulations at different pressures (from left to right: 25, 50, 150, 250, 350 GPa).\\
}
\label{fig:pd}
\end{figure*}

Liquid hydrogen constitutes the interior of giant planets and brown dwarf stars,
and it is commonly assumed to undergo a first-order phase transition between an insulating molecular fluid and a conducting metallic fluid\cite{mcmahon_properties_2012}.
Understanding the nature of this liquid-liquid transition (LLT) is crucial to accurately model the structure and evolution of giant planets including Jupiter and Saturn\cite{guillot_interiors_2005}.
Standard planetary models assume a sharp LLT accompanied by a discontinuity in density, and as a result feature a clear-cut transition between an inner metallic mantle and an outer insulating mantle\cite{Hubbard_2016}.
Probing the nature of the LLT in laboratories faces the challenges of creating a controllable high pressure and temperature environment, and of confining hydrogen whilst making measurements.
As such, experimental studies have not yet reached a consensus on whether the LLT is first-order or smooth\cite{Celliers677},
and there are considerable discrepancies up to 100 GPa (see Figure~\ref{fig:pd}) on the location of the LLT 
between experiments\cite{knudson2015direct,Ohta2015,McWilliams2016,Zaghoo2016,Zaghoo2017,Celliers677}.

Given the experimental difficulties, computer simulations have played a fundamental role in characterizing the phase diagram of hydrogen\cite{scandolo2003liquid,morales_evidence_2010,lorenzen,Bonev2004}, by describing atomic interactions based on a quantum-mechanical treatment of electrons. 
Different levels of electronic structure theory have been employed, ranging from accurate Quantum Monte Carlo (QMC) methods\cite{delaney2006quantum,morales_evidence_2010,PhysRevLett.120.025701}, to the popular Density Functional Theory (DFT) approximation\cite{scandolo2003liquid,Vorberger2007,morales_evidence_2010,lorenzen}.
Early simulations gave contradictory results~\cite{scandolo2003liquid,delaney2006quantum,Vorberger2007}, but most recent calculations identify small density discontinuities at below 1500~K \cite{morales_evidence_2010,lorenzen,PhysRevLett.120.025701}, which were interpreted as signatures of a first-order LLT.

Even at the lower computational cost end of electronic structure methods, DFT studies of dense liquid hydrogen are limited to a size of few hundreds atoms and a time scale of a few picoseconds (ps)\cite{scandolo2003liquid,Vorberger2007,morales_evidence_2010,lorenzen}. Given the subtlety of the problem, it would be desirable to overcome these size and timescale limitations.
To this end, we use an artificial neural network architecture to construct a machine-learning potential (MLP) (see details and benchmarks in the \si),
based on the Behler-Parrinello framework~\cite{behler2007generalized}.
We have tested extensively the MLP against direct ab initio simulations on small systems, observing excellent agreement. 
The combination of the first-principles accuracy and the low cost of the MLP
allows us to investigate hydrogen phase transitions for temperatures ($T$) between 100 and 4000~K, and pressures ($P$) between 25 and 400~GPa, with converged simulation size and time.
If performed using DFT, the total computational cost of this study would require several millions of CPU years, exceeding the capacity of the world's fastest supercomputers.

\paragraph*{Solid-liquid transition}

Computing the hydrogen melting line is nontrivial because solid hydrogen exhibits polymorphism and only a few of the crystal structures and phase boundaries have been characterized conclusively~\cite{mcmahon_properties_2012,howie_mixed_2012,zha_high-pressure_2013,dalladay2016evidence}.
Given that estimating the melting point using most free energy methods requires a prior knowledge of the stable crystal structure under each pressure, we simulated the cooling of 1728-atom hydrogen systems from well-equilibrated liquid to frozen structures, 
and subsequently re-heated them until melting, for a total of 1.6 ns of simulation time at each pressure. Due to hysteresis, the freezing and melting temperatures bracket the melting point $T_m$, that can be estimated as the mean of the two values, as indicated by the black curve in Fig. \ref{fig:pd}a.
The shape of the melting line, with $T_m$ peaking around 100 GPa and 1000~K, followed by a decline at higher P, are consistent with recent  experimental measurements~\cite{zha2017melting}.

The solid configurations obtained at different pressures (Fig.~\ref{fig:pd}c) follow a trend consistent with both experimental evidence~\cite{mcmahon_properties_2012,howie_mixed_2012,zha_high-pressure_2013,dalladay2016evidence} and previous first principles simulations~\cite{monserrat2016hexagonal}, going from close-packed structure with freely rotating molecules at low $P$  to aligned and layered structures at higher pressure.
At $P \approx 350$~GPa we observe the transition into a novel atomic phase, with molecules dissociating to form long wires arranged in a hexagonal lattice. 
We provide the structures of the solid phases in the \si, which can be compared with the
crystal structure predictions by random search using small simulation cells~\cite{Needs2016},
and be used to characterize the phase diagram of dense solid hydrogen.

\paragraph*{Liquid-liquid transition}

After having located the melting line for the MLP, we performed equilibrium molecular dynamics simulations across a broad range of $T$ and $P$ in the liquid region, 
with simulation time and system size sufficient to achieve convergence.
We computed the fraction of molecular hydrogen by counting the H--H bonds using a smooth cutoff function that decays from one to zero between 0.8~\AA{} and  1.1~\AA{}.
As evident from Fig.~\ref{fig:pd}a, the molecular fraction varies smoothly across the liquid phase diagram, with the transition region becoming narrower at low $T$ and high $P$. 
Other observables, including density ($\rho$) and heat capacity $c_P$ (Fig.~\ref{fig:pd}c), the pair correlation function and the electronic density of states (see \si), all similarly show no sign of discontinuities. 
Both $\rho$ and $c_P$ exhibit an anomalous behavior, displaying smooth peaks that become sharper at higher pressures (Fig. ~\ref{fig:pd}b) 
The loci of these  maxima, as well as the atomic-molecular transition region, converge towards the melting line around 350 GPa.

The results above rule out a first-order LLT for the hydrogen system described by the MLP.
However, as discussed in the literature~\cite{lorenzen,morales_evidence_2010}, the location of the boundary of the LLT is sensitive to the details of the underlying electronic structure methods, which could also affect the melting line and the solid phase diagram.

\paragraph*{Polyamorphic solution model}

\begin{figure*}[hbt]
\includegraphics[width=0.8\textwidth]{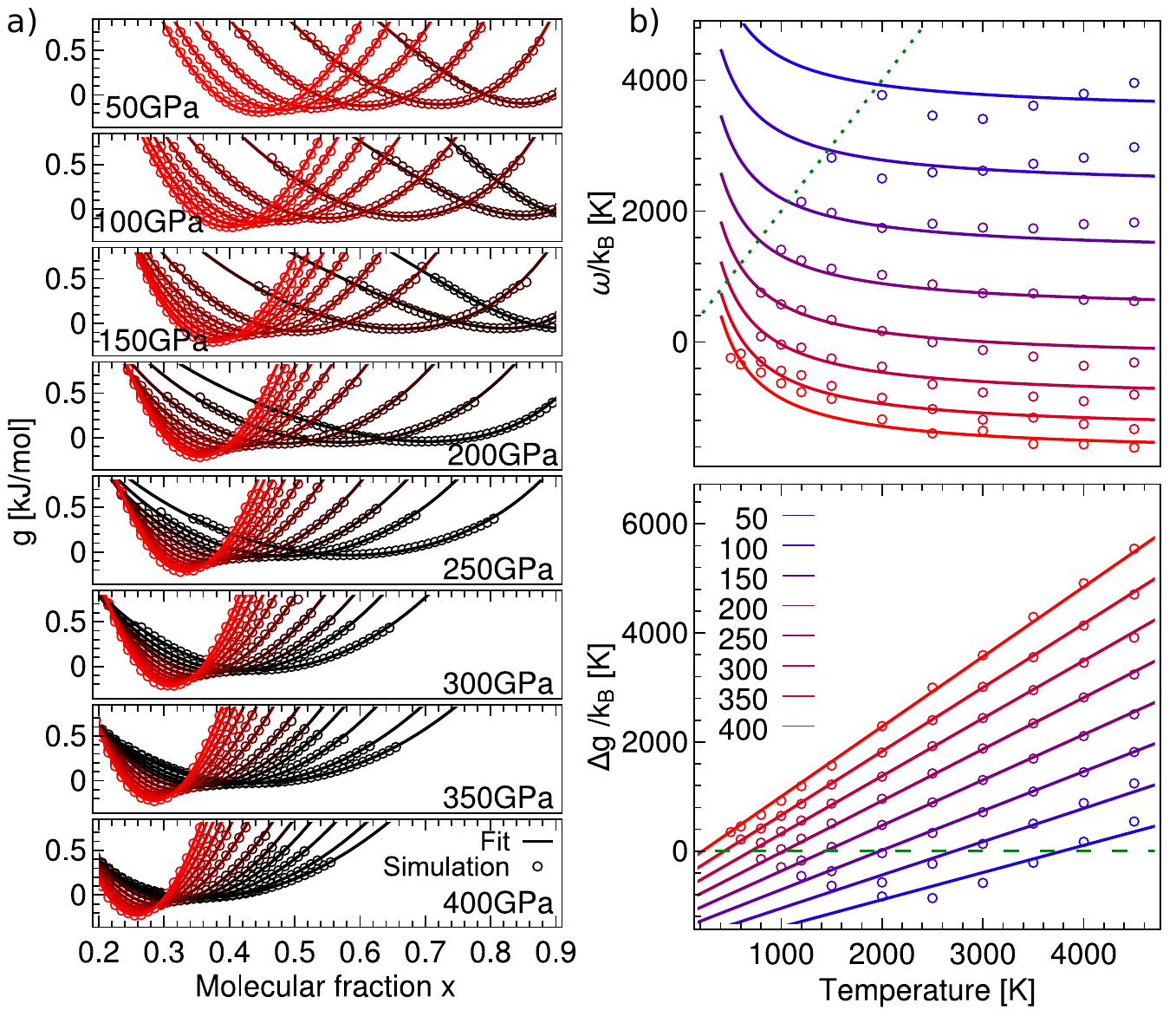}
\caption{
Panel (a): The dots show the computed free energy profiles $g(x)$ as function of fraction of molecular hydrogen at different constant pressure and temperature conditions,
and the smooth curves show the individual fits to the polyamorphic solution model.
From dark to red color, results at temperatures of 500, 600, 800, 1000, 1200, 1500, 2000, 2500, 3000, 3500, 4000, 4500K are plotted.
$g(x)$ at higher temperature has a lower molecular fraction.
Panel (b): 
On the lower panel, the dots are the fitted values of $\Delta g = g_M-g_A$ to the solution model,
and the lines are linear fits to the values of $\Delta g$.
On the upper panel, the dots are the individual values of $\omega$ obtained from fitting $g(x)$ to the solution model at different $P$ and $T$,
and the curves are the fits to those values.
The dotted green line corresponds to $\omega=2T$, that corresponds to the phase separation line, and the dashed line to $\Delta g=0$, i.e. the coexistence line. 
}
\label{fig:solution}
\end{figure*}

To provide a more robust analysis of the nature of the LLT, and to propose a thermodynamic model that can be used to interpret experimental observations, we map our simulations on a polyamorphic solution model~\cite{anisimov2018thermodynamics} that describes a mixture of two inter-convertible liquid states.
At each thermodynamic state point, the regular-solution molar free energy $g(x)$ as a function of the molecular fraction $x$ reads
\begin{equation}
g(x) = x \Delta g +k_B T x \ln(x)
    +k_B T (1-x)\ln(1-x) + \omega x(1-x)
    \label{eq:solution}.
\end{equation}
The term $\Delta g = g_M-g_A$ is the chemical potential difference between the atomic and molecular phases, and $\omega$ is an enthalpic term that accounts for the non-ideality of mixing.
In order to obtain a free-energy profile from simulations, that can be compared to Eq.~(\ref{eq:solution}), we performed a separate set of calculations in which we enhanced the spontaneous fluctuations of the order parameter $x$ using metadynamics~\cite{laio2002escaping} (see \si). 
As shown in Fig.~\ref{fig:solution}a, the model matches perfectly the $g(x)$ obtained from simulations, that exhibits a single minimum, indicative of perfect mixing of the two liquids and of the absence of a LLT throughout the range of temperatures and pressures that we explored. 
In addition, the fitting parameters are well described by the simple empirical models
$\Delta g = a_0 + a_1 P + a_2 T + a_3 P T$,
and
$ \omega = b_0 + b_1 P + b_2/T + b_3 P^2$. As shown in Fig.~\ref{fig:solution}b, the global model matches well the values obtained by independent fits at each state point. 
These analytic expressions make it possible to estimate the $x=0.5$ coexistence line  $T_{0.5}$ (i.e. the temperature at which atomic and molecular fluids would be equally stable, determined implicitly by $\Delta g(P,T)=0)$),
as well as the phase separation line $T_s$ (i.e. the temperature below which the two fluids start de-mixing, determined by $T_s=\omega(P,T_s)/2k_B$). 
There are several considerations one should keep in mind when discussing this model. (1) 
There are different ways to define the molecular fraction $x$ using a local order parameter,
however, the curves $T_s$ and $T_{0.5}$ are rather insensitive to such definition. (2) Given that the two phases can interconvert, being at $T<T_s$ is not sufficient to observe two-phase behavior - it is also necessary that the atomic and molecular phases are equally stable. (3) The model describes bulk thermodynamics, and as such the estimation of $g(x)$ from biased atomistic simulations is only possible for $T>T_s$, since an artificially-stabilized two-phase configuration would also involve a free-energy penalty associated with the phase boundary.
The two temperatures are plotted on Fig.~\ref{fig:pd}a as dashed and dotted green lines and cross at the critical point marked by a green star for the fluid-fluid phase transition, that is located at $(P_c,T_c)\approx$350~GPa and 380~K, which coincides approximately with the melting line, and with the point at which atomic solid phases appear.
At $T>T_c$ the system exhibits supercritical behavior, without phase separation and with anomalies in the thermodynamic properties of the mixture following different Widom lines that emanate from the critical point. At $T<T_c$ the system exhibits a first-order phase transition, and $T_{0.5}$ determines the coexistence line between the two phases. 
Since in this system $T_c<T_m$, no LLT can be observed, even though the anomalous behaviors induced by a hidden critical point can be observed throughout the liquid phase diagram, much like the case of water~\cite{soper2000structures}.

Our observation of a supercritical fluid above the melting line contradicts several recent DFT and QMC simulations\cite{lorenzen,morales_evidence_2010,PhysRevLett.120.025701}, which reported a sharp LLT suggested by small discontinuities in density up to around 1000-1500 K.
We discuss in detail the probable origin of this discrepancy in the \si{}, which we traced to finite-size effects resulting in the formation of a highly-defective solid phase.
By following the simulation protocol of previous studies\cite{lorenzen}, we could reproduce the density discontinuities employing either DFT or the MLP at $T\le1000K$.
This transition is associated with a sharp drop of diffusivity, and with the appearance of ordered planes of molecules. Repeating simulations in the isothermal-isobaric ensemble leads to effectively zero diffusivity, and we observe the appearance of a close-packed solid phase similar to the configurations obtained by slow quenching using the MLP, up to $T=1250$~K.

We note that, in general,
a change in electronic structure method, or the inclusion of nuclear quantum effects,
or the residual difference between MLP and the underlying first principle potential energy surface,
can change quantitatively the predicted phase diagram.
In fact, the use of MLP should facilitate greatly the comparison between different electronic structure methods, possibly also allowing to obtain thermodynamic observables free of residual inaccuracies introduced by ML schemes using free-energy perturbations~\cite{chen+19pnas}.
Nevertheless,
the polyamorphic solution model brings robustness to the qualitative prediction of supercritical behavior:
at $P<200$ Gpa, $T_{0.5}$ is well above $T_s$, meaning that the atomic fluid is unstable at conditions where phase separation can happen.
At about $P>200$ Gpa, $\omega$ is negative at $T\gtrsim 800~K$, suggesting that mixing is enthalpically - and not only entropically - favorable.

The predicted supercritical behaviors of fluid hydrogen
could explain the discrepancies between different experiments.
While all observables should undergo an abrupt change when crossing the coexistence line 
in the case of a first-order LLT, 
the supercriticality of fluid hydrogen means that the boundary of LL transition is blurry and its location depends on the specific criterion used to define it.
In other words, different
observables may exhibit an anomalous behavior that follows different Widom lines, as we observed for density and heat capacity.
Indeed, the LLT boundaries observed by different teams at $ 1000 < T < 2000$ K all qualitatively extrapolate toward the proposes critical point (see figure 1 a)\cite{knudson2015direct,Ohta2015,McWilliams2016,Zaghoo2016,Zaghoo2017,Celliers677}. The observation of a sharper transition in the low-temperature compression experiments of Knudson et al.~\cite{knudson2015direct}, in comparison to those performed by Cellier et al.~\cite{Celliers677} at higher $T$, is also consistent with supercritical behavior.
We propose that a polyamorphic solution framework, which we validated in our simulations, could be used to macroscopically model the stability and miscibility of atomic and molecular hydrogen.
Such a model provides a thermodynamic understanding of the LLT to directly interpret experiments and astrophysical observations in a way that accounts for the presence of a molecular-to-atomic transition in dense liquid hydrogen, with evidence against the existence of a first-order transition. 
Our approach, combining machine learning potentials trained on electronic structure calculations, thorough statistical sampling and macroscopic thermodynamic models, can be used to quantitatively assess the properties of mixtures of hydrogen and heavier elements, to address long-standing questions concerning Jupiter's core solubility and the anomalous luminosity problem of Saturn\cite{guillot_interiors_2005}.

\clearpage
\ \vfill
{\centering \Huge Supporting Information\\}
\vfill
\foreach \x in {2,3,4,5,6,7,8,9,10,11,12,13,14,15,16,17,18,19,20,21,22}
{%
\clearpage


\section*{Supplementary material}

# Supporting Information

# CONTENTS

# I. DETAILS OF PBE DFT CALCULATIONS

For generating the training set of the machine learning interatomic potential, as well as for running reference *ab initio* molecular dynamics simulations, we employed density functional theory (DFT) using the PBE approximation to the exchange-correlation energy. For all the DFT calculations performed in this study, we used a cubic simulation cell of 128 H atoms, $4 \times 4 \times 4$ Monkhorst-Pack k-points, 80 Rydberg (Ry) plane-wave cutoff, and the PAW Pseudopotential [1]. The QuantumEspresso 6.2.0 package [2] was used and all the input files are provided as a part of the Supplemental Information.

# II. *AB INITIO* MOLECULAR DYNAMICS SIMULATIONS USING DFT AT CONSTANT VOLUME AND TEMPERATURE ($NVT$)

## A. $NVT$ AIMD simulation details

We performed *ab initio* molecular dynamics (AIMD) simulations using PBE DFT (see Sec I) at the constant volume and temperature ($NVT$) ensemble for a system of 128 hydrogen atoms, over a broad range of densities. The Wigner-Seitz radius $r_s$ is the radius of a sphere whose volume is equal to the volume per atom in the units of the Bohr radius. We express density in terms of $r_s$, as it is often done in the context of studies of high-pressure hydrogen. The density $\rho$ described by the Wigner-Seitz radius (which is used in other works) can be converted to g/mL via the relationship $\rho[\text{g/mL}] = 2.6966/(\rho[r_s])^3$. For the AIMD simulations, we considered densities ranging from $r_s = 1.26$ to $r_s = 1.60$ (i.e. 1.348-0.658 g/mL), as well as at a broad range of temperatures (600 K, 800 K, 1000 K, 1200 K, 1500 K, 2000 K, 3000 K, 6000 K, 8000 K). We used a time step of 0.2 fs, and a loose convergence criterion of $10^{-4}$ Ry for the self-consistence loop. We used a strong stochastic velocity rescaling thermostat [3], with a time constant of 10 fs, to compensate for the relatively high level of noise on the forces, following the idea of Ref. 4. The simulation time for each $NVT$ simulation at each density and temperature is about 0.8 ps, and we allow for 0.4 ps of equilibration time before computing the system properties. The starting atomic structures of the AIMD simulations were taken from previous QMC simulations[5] at the corresponding density.

## B. *NVT* AIMD simulation results

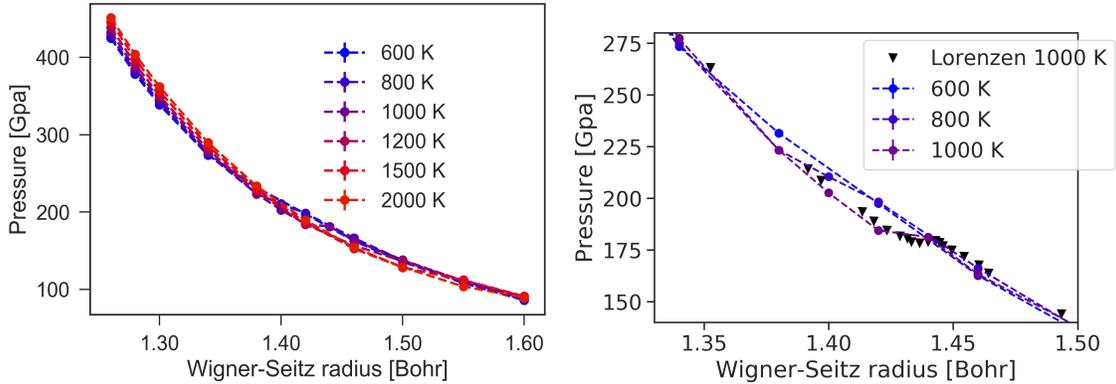

FIG. S1: Pressure at each temperature and density. Results computed from AIMD simulations using PBE DFT with a system of 128 H atoms. The right panel is a zoomed-in figure, to show the discontinuities in the pressure-density relations for the 800 and 1000 K isotherms, and to compare with available AIMD results from Ref. 6 using PBE functional.

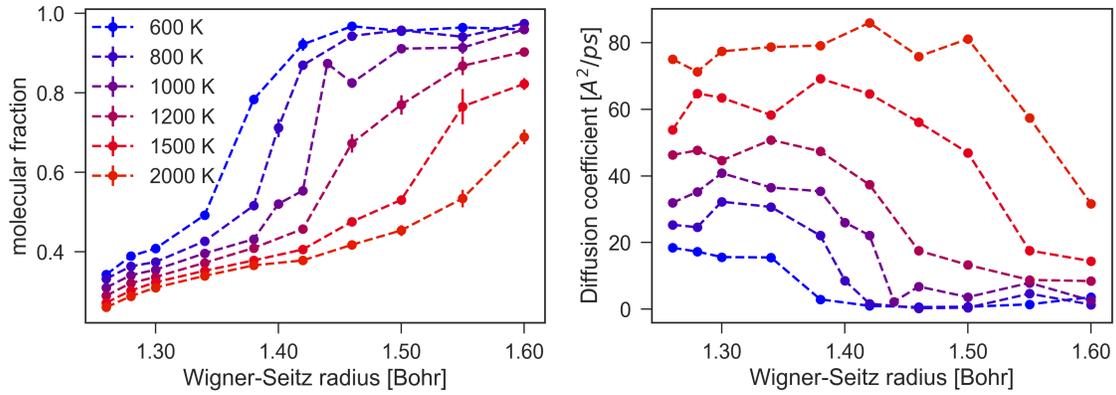

FIG. S2: Left: Fraction of molecular hydrogen atoms. Right: Diffusion coefficients. Results computed from AIMD simulations using PBE DFT with a system of 128 H atoms.

To trace the possible phase transition, and to compare with previous DFT calculations [6], we computed the average pressure for each *NVT* AIMD simulations at each density and temperature (Fig. S1 left panel). With the density grid that we used, the $P - \rho$ curve at isothermal conditions with $T \geq 1200$ K appears smooth, with small jumps that can be observed at lower temperatures (1000 K and 800 K). This is consistent with previous DFT findings [6, 7].

4

Much stronger signals on the atomic-molecular transitions are revealed by calculating the fraction of bonded atoms in the system (Fig. S2 left panel), defined as bonded atoms with one neighbor within a smooth cutoff that starts from 0.8Å and decays to zero at 1.1Å, as well as the diffusion coefficient plotted on the right panel. The molecular to atomic transition is also captured by the H-H radial distribution functions (RDFs) and vibrational density of states (VDOS) plotted in the left panels of Fig. S9 and S10, respectively.

### C. Discussions on the nature of the phase transition

Although the molecular to atomic transition as the density increases is evident from these simulation results, it is difficult to interpret if the transition between atomic and molecular states is smooth or abrupt, especially considering that the small system size (i.e. 128 atoms) may affect the diffusivity as well as phase stability [8]. In addition, the simulation time of about 0.8 ps is very short, making it hard to ascertain whether the system has reached equilibrium.

Another problem is that solidification may happen during the AIMD simulations and the system becomes molecular solid hydrogen as a result. Given that the transition temperatures associated with the discontinuities are very close to the melting line identified in slow quenches of the machine-learning potential (MLP), one should consider carefully whether the observed discontinuity could be associated to solidification, rather than to a LLT. Solid hydrogen has a rich phase diagram, and in some phase molecules have rotational degree of freedom, which makes the classification of the crystal structure difficult. When solidification does happen in the AIMD simulations, because of the small system size and the constant volume condition, the solid can carry a high amount of defects, which further complicates the identification process. Indeed, some structures obtained from our AIMD simulations seem ambiguous, and we found it difficult to decide if they are crystalline. Nevertheless, many molecular structures obtained from AIMD simulations have clear crystalline-like features. As shown in Fig. S3, at 1000 K, 800 K or 600 K, the snapshots generated at certain densities have hydrogen molecules arranged on lattice, which highly resemble molecular solid hydrogen structures. Indeed, along these isotherms, whenever the molecular fraction (Fig. S2 left panel) reaches $\approx 0.8$ or the diffusion coefficient (Fig. S2 right panel) drops to $\approx 5$Å/ps, the crystalline features become salient.

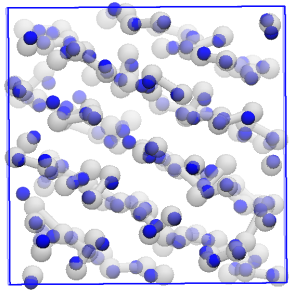

(a) $T = 600$K, $r_s$ 1.38 Bohr

($\langle P \rangle = 230$ GPa)

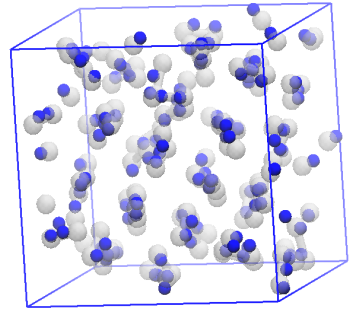

(b) $T = 800$K, $r_s$ 1.42 Bohr

($\langle P \rangle = 200$ GPa)

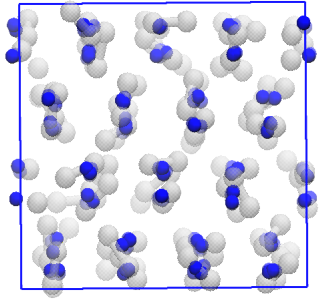

(c) $T = 800$K, $r_s$ 1.46 Bohr

($\langle P \rangle = 170$ GPa).

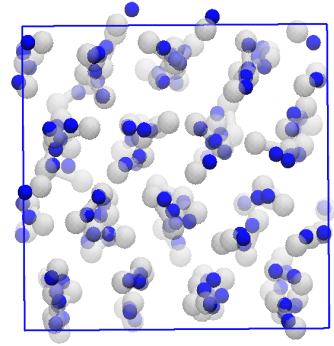

(d) $T = 800$K, $r_s$ 1.46 Bohr

($\langle P \rangle = 170$ GPa).

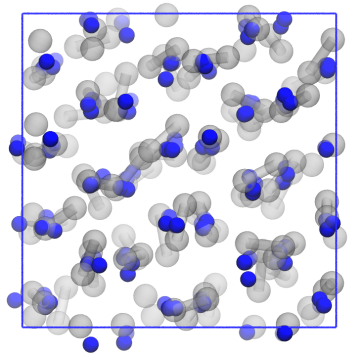

(e) $T = 1000$K, $r_s$ 1.44 Bohr

($\langle P \rangle = 180$ GPa)

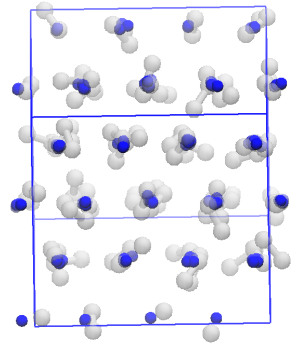

(f) $T = 1000$K, $r_s$ 1.60 Bohr

($\langle P \rangle = 90$ GPa).

FIG. S3: Snapshots from AIMD simulations at $NVT$ ensemble. Blue dots correspond to the positions of the H atoms averaged over a 200 fs window, corresponding roughly to the center of mass of rotating molecules. (c) and (d) are taken from the same simulation but at different time steps.

16

In particular, the presence of a discontinuity in the pressure-density curve is most apparent at $T = 1000$ K, similar to what has been observed in Ref. [6], and so we focus our attention on the region of the phase with $T = 1000$ K and $r_s \approx 1.42$ Bohr. As shown in Fig. S2 (right panel) the transition is associated with a decrease in diffusion coefficient, that at low temperature becomes very small as the density is decreased. Indeed, an inspection of the MD trajectory reveals the presence of layers of molecules, which is reminiscent of some of the solid H phases we observed in the MLP solid (Fig. S3 (e)).

To sum up, we observed molecular-atomic transitions of high pressure hydrogen in constant-volume AIMD simulations at different densities, which are consistent with previous results. At low temperatures ($T \leq 1000K$), such transitions seem rather sharp, but an inspection on the snapshots of atomic coordinates reveals that hydrogen has frozen into a molecular solid during the transitions. When liquid hydrogen freezes into a molecular solid, the P-$\rho$, molecular fraction and diffusivity curves will show a sharp change. At higher temperatures ($T \geq 1200K$), the transitions are smoother compared with the ones at lower temperatures, the atomic structures are not obviously similar to crystals, and the nature of the transition seem ambiguous.

## III.   *AB INITIO* MOLECULAR DYNAMICS SIMULATIONS USING DFT AT CONSTANT PRESSURE AND TEMPERATURE ($NPT$)

### A.   $NPT$ AIMD simulation details

As discussed in the previous section, the nature of the transition can be obfuscated by the small system size and the constant-volume constraint. To obtain a clearer picture and to investigate the effect of the $NVT$ conditions, we conducted isothermal-isobaric ($NPT$) simulations over a range of pressures. The details of the DFT calculations are identical (see Sec. I), and the simulation cell contains 128 hydrogen atoms. Langevin piston barostat [9] was employed to maintain constant pressure conditions. To compensate for the slow dynamics of the simulation cell, and to increase the quality of statistical averages, we extended the simulation time to about 2 ps, allowing for 0.4 ps of equilibration time before computing the system properties.

## B.  *NPT* AIMD simulation results, and comparison with *NVT*

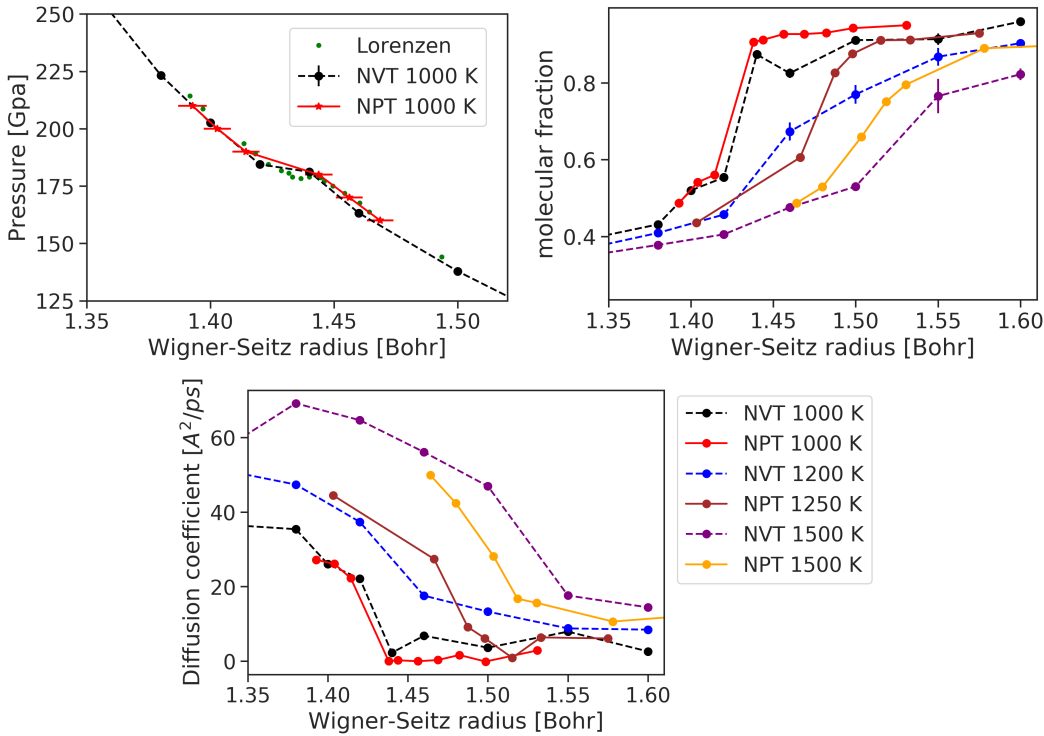

FIG. S4: Upper left: Pressure at each density. Results from Ref. [6] at $T = 1000K$ are also shown. Upper right: Fraction of molecular hydrogen atoms. Bottom: Diffusion coefficients. Results computed from AIMD simulations using PBE DFT with a system of 128 H atoms at the $NVT$ or the $NPT$ ensemble.

From the AIMD simulations at $NVT$, we observed quite sharp transitions at low temperatures $T \leq 1000K$ (Fig. S2), which has been shown to be related to crystal formation (Fig. S3). At 1000 K, the transition is more clear-cut in isobaric conditions, and the diffusion coefficient drops effectively to zero (see Figure S4), suggesting that constant-volume constraints increase the concentration of defects. Snapshots taken from the equilibrated part of the trajectory reveal unambiguously the formation of a solid phase, with freely-rotating molecules whose center of mass lie on average on a close-packed lattice (Fig. S5 (d)). At higher temperature, however, the picture from $NVT$ simulations seemed less clear. To probe the nature of the transition under these conditions further, we also performed AIMD $NPT$ simulations at 1250 K, and 1500K. The pressure-density curve, fraction of molecular hydrogen and diffusivities corresponds roughly to that computed in the $NVT$ ensemble. At

8

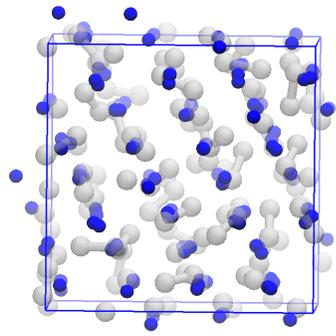

(a) $T = 1250$K, $P = 130$ GPa

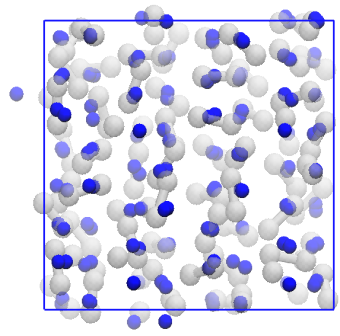

(b) $T = 1000$K, $P = 140$ GPa

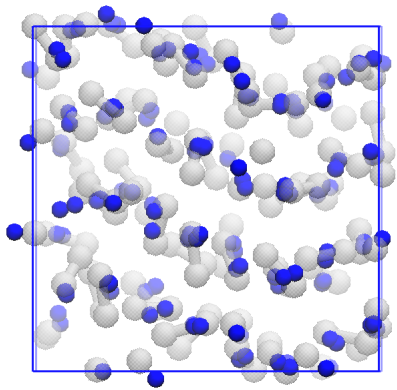

(c) $T = 1000$K, $P = 160$ GPa

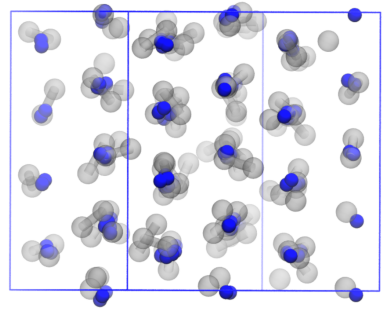

(d) $T = 1000$K, $P = 180$ GPa

FIG. S5: Snapshots from AIMD simulations at $NPT$ ensemble. Blue dots correspond to the positions of the H atoms averaged over a 200 fs window, corresponding roughly to the center of mass of rotating molecules.

pressures from 120 GPa to 180 Gpa, we observed the formation of clear crystalline-like structures from AIMD simulations at temperatures up to 1250 K, (Fig. S5), providing further evidence in support of the solid-liquid nature of the transition observed in this region of the phase diagram. The diversity, and the high concentration of defects observed for these structures may be related to the polymorphism of the solid H diagram, and underscores the fact that the study of the solid-liquid phase boundary requires a complete understanding of the solid-state portion of the phase diagram.

As discussed in Sec. V B, phase transition behaviors in a small system size suffer from finite size effects, which leads to the broadening and the smoothing of the transition region.

The very occurrence of crystallization in the AIMD simulations within the time scale of picoseconds is, in fact, thanks to such finite size effects. As such, to probe the phase behavior of the system at the thermodynamic limit, it is essential to employ large system size, which, of course, can become prohibitive using DFT.

## IV. MACHINE-LEARNING POTENTIAL FOR HIGH PRESSURE HYDROGEN

### A. Training and benchmark

We constructed a neural-network-based machine-learning potential (MLP) for bulk H at high pressure, which was trained based on PBE energies and forces for 9,194 diverse reference structures of 128 atoms of H. The setup of the DFT calculations was identical with the one described in Sec I. Among the training set, 5,058 configurations were selected from a previous data set [5], and in order to describe well the dense hydrogen systems at pressure $P > 250 GPa$, another 4,136 configurations were selected from a part of the AIMD simulation trajectories at high pressure generated in the present work.

Based on this training set, we generated a flexible and dissociable machine learning potential for high-pressure hydrogen, employing an artificial neural network architecture built according to the framework of Behler and Parrinello [10]. The training was performed using the N2P2 code [11]. Within this framework, the total energy of the system is expressed as the sum of the individual contributions from atom-centered environments, which encompasses the relative coordinates of all neighboring atoms inside a cutoff radius. In the case of high pressure hydrogen, we have systematically performed convergence tests employing cutoffs ranging from 6-12 Bohr, and selected a cutoff of 8 Bohr. To remove rotational as well as permutation variances of atomic coordinates, we selected a total of 84 Behler-Parrinello symmetry functions (SFs) to describe the atomic environments according to the correlation between the values of the SFs and the magnitudes of forces on central atoms. The values of the SFs are then used as input vectors for the atomic neural networks, which contain two hidden layers with 20 nodes each, yielding the atomic energy contributions. Finally, the analytic total energy expression is a sum over the outputs of all individual atomic neural networks, and analytic gradients for the calculation of the forces are readily available.

The resulting root mean squared errors (RMSE) of the energies in the training and the

test set are both 5 meV/atom, while the RMSE values of the forces in both sets are 300 meV/angstrom.

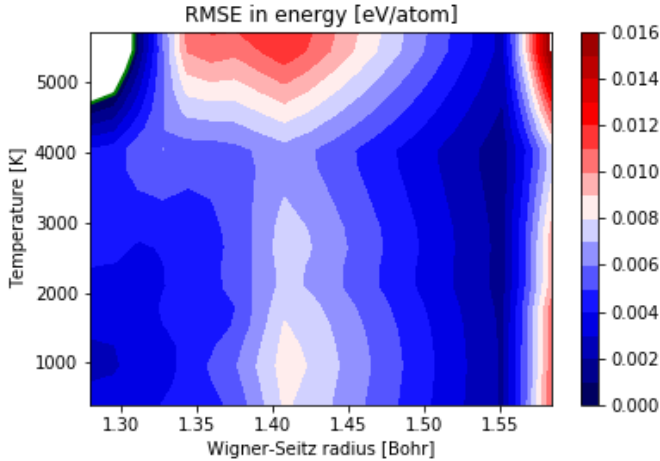

FIG. S6: The root mean squared errors (RMSE) of the energies between between MLP and PBE DFT for more than 200,000 configurations generated from AIMD simulations at different temperatures and densities.

In order to benchmark the MLP across all the thermodynamic conditions relevant to this study, we took all the trajectories from the AIMD simulations, and compared the predictions of the potential energy of the H systems between MLP and PBE DFT for these confiurations. From Fig. S6, it can be seen that the root mean squared errors (RMSE) of the MLP energies are below 10 meV/atom for temperatures below 6000 K and densities between 0.658-1.348 g/mL, which covers the pressure range of about 50 GPa to 450 GPa . Note that only 4,136 configurations in the training set of the MLP were selected from more than 200,000 AIMD configurations, and most AIMD runs were not included in the training, so that several thermodynamic conditions represent genuine predictions.

### B.  Validation of the neural network potential and comparison with DFT

For further validation, we performed $NVT$ simulations for systems of 128 H atoms using the neural network potential of hydrogen, so as to compare to the results from ab initio molecular dynamics simulations using DFT with PBE functional approximations described in Section II A. The $NVT$ simulations using the MLP have been carried out using

11

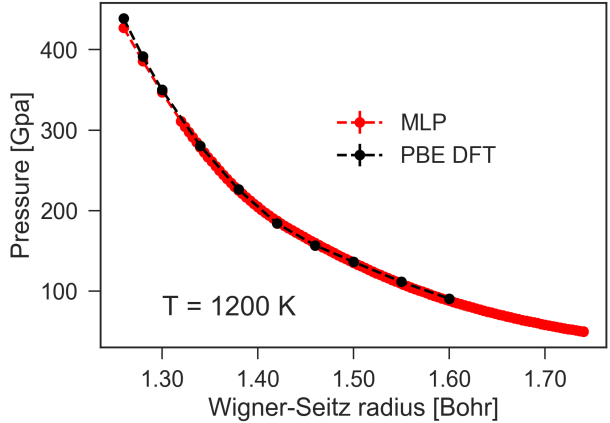

FIG. S7: A comparison between the equilibrium pressure at 1200 K and a range of densities predicted by PBE DFT and the machine-learning potential.

LAMMPS [12] patched with N2P2 [11]. The simulations were performed for a range of temperatures from 600K to 6000 K, and for a range of density corresponds to 1.26 - 1.74 Wigner-Seitz radius $r_s$, corresponding to a pressure range of 50 GPa - 400 GPa. For each temperature and density, we ran two independent simulations either starting from an atomic or a molecular configuration, and the simulation time for each run is about 400 ps. With such simulation length, we were able to ensure that both simulations with different starting configurations give converged results.

In Fig. S7, we show the comparison between the MLP simulations and AIMD for the $P - \rho$ curve at 1200 K isothermal conditions. We also compared the fraction of molecular hydrogen and the diffusion coefficient at different densities and temperatures (see Fig. S8). In general, the results from the MLP show good agreement in all aspects. In particular, the slope of change of molecular fraction as density increases is very well captured by the MLP compared with DFT. The transition exhibits jumps at $T < 1000$ K temperatures, that are associated to solidification, but appears smoother at higher $T$. The MLP seems to slightly over-shoot the transition density by about $r_s = 0.03$ (0.05 g/mL in SI units), which roughly corresponds to an overestimate in the transition pressure by about 25 GPa. In general, the MLP seems to predict a smoother transition than DFT in the ambiguous region, with 128-atoms simulation boxes. This may be due to an underestimation of the melting point, or to an underestimation of the solid-liquid interfacial energy, that makes the formation of defects and fluctuations between the two phases more facile. In all cases, the MLP seems to capture

12

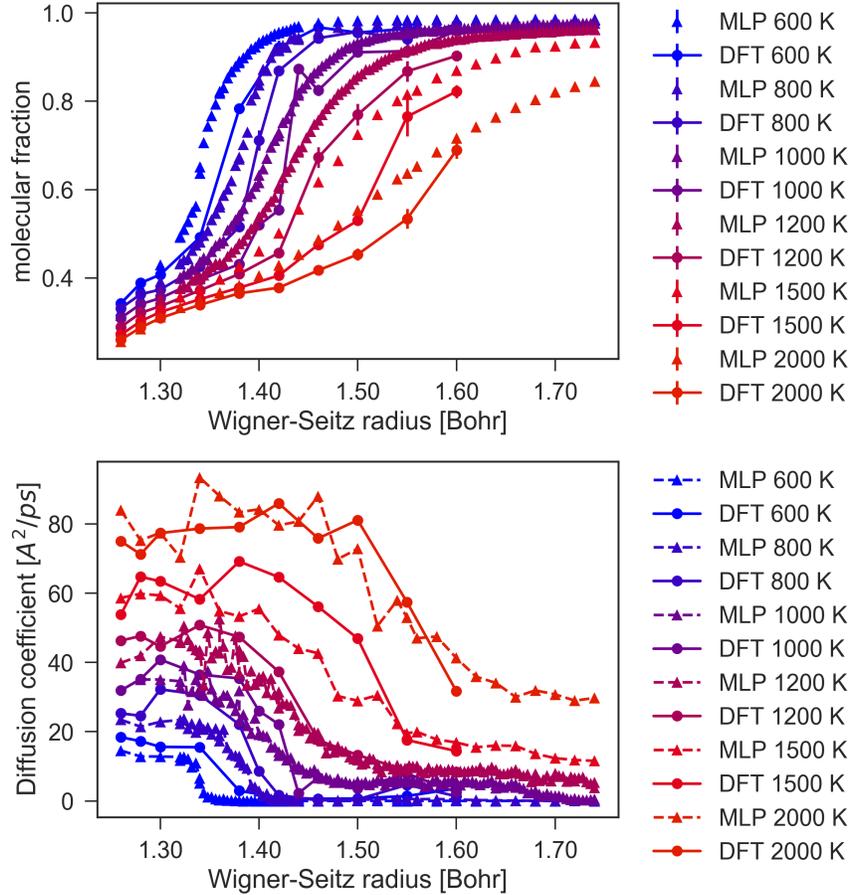

FIG. S8: A comparison between predictions from PBE DFT and the machine-learning potential at different temperatures and densities. Upper: Fraction of molecular hydrogen atoms. Lower: Diffusion coefficients.

well the qualitative features of the transition, making it well-suited to the systematic study of finite-size effects that we discuss in Sec. V A.

Furthermore, we have compared the H-H radial distribution functions (RDFs) and vibrational density of states (VDOS), as shown in Fig. S9 and S10, respectively. Across all the temperatures and densities considered there is a good agreement between the MLP and the DFT description. The agreement is particularly good when the H system is mostly atomic or mostly molecular, as near the LLT, the MLP displays a slight shift of the transition density compared with the DFT. At high temperatures the agreement is better, probably because that at low temperatures the small systems sometimes freeze into solids with defects and deformations under the $NVT$ ensemble, as shown above. The random nature of the residual defects affects the reliability of this comparison in the vicinity of the melting line. Notice

13

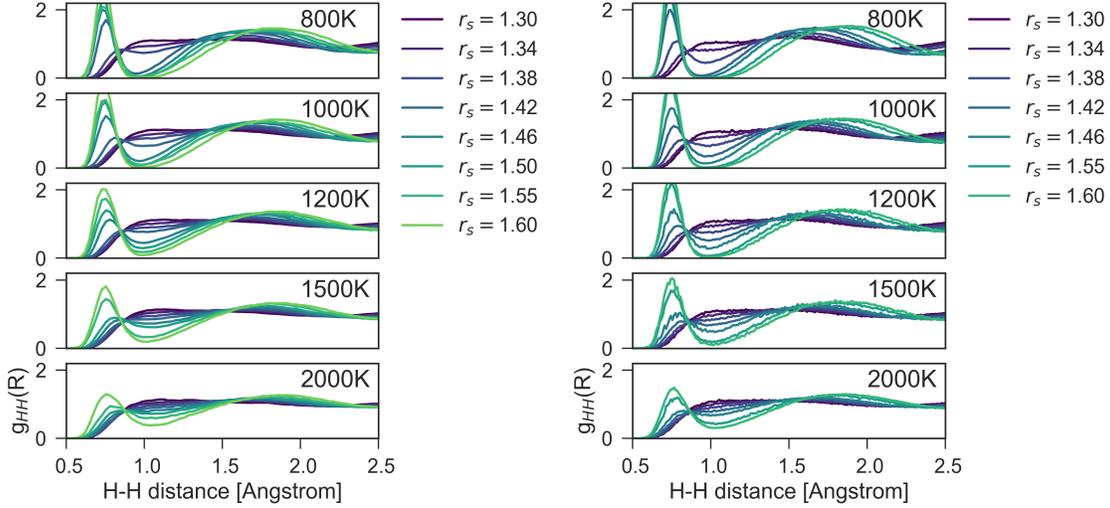

FIG. S9: H-H radial distribution functions $g_{HH}$ computed from equilibrium molecular dynamics simulations at $NVT$ ensemble with a system size of 128 atoms. Panel (a): using PBE DFT. Panel (b): using ML potential.

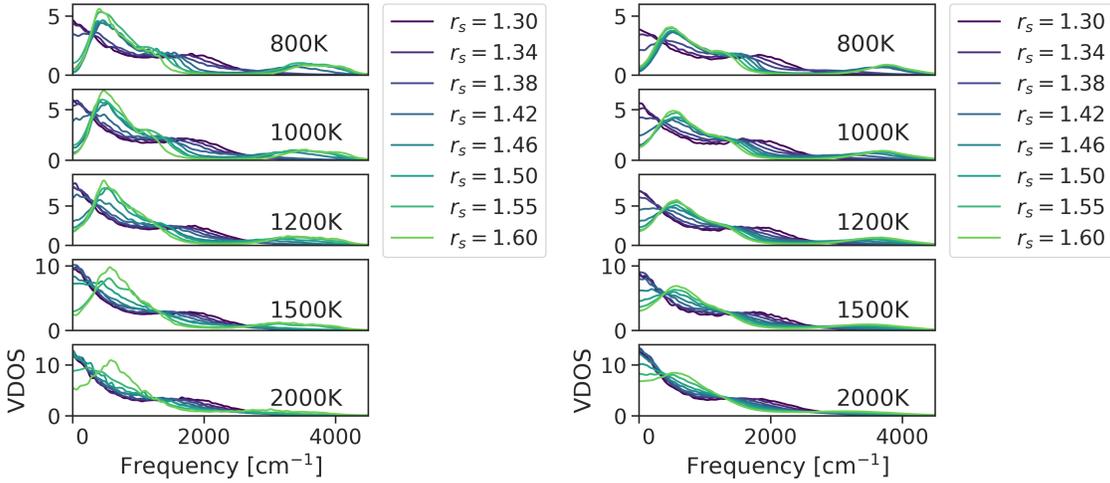

FIG. S10: Vibrational density of states (VDOS) of H atoms computed from equilibrium molecular dynamics simulations at $NVT$ ensemble with a system size of 128 atoms. Panel (a): using PBE DFT. Panel (b): using ML potential.

also that there seem to be an isosbestic point in the RDFs at about 0.9 Å, and in the VDOS plots at about the frequency of 1200 cm$^{-1}$ at 1500 K and 2000 K, which suggests the system behaves like a mixture of two species.

14

## V. FINITE SIZE EFFECTS

### A. Finite size effects in $NVT$ simulations

In previous DFT and QMC simulations [5–7], the demonstration of a first-order liquid-liquid transition is mostly based on a plateau in pressure (P) versus density ($\rho$) along isotherms for a small hydrogen system, using a simulation setup and system size similar with the AIMD simulations described in Sec II A, that involves constant-volume isothermal simulations. In fact, it appears that the discontinuity corresponds to solidification, that is obscured in constant-volume simulations due to the high concentrations of defects, while simulations in the isobaric ensemble exhibit clearer crystalline order.

The difference between the two ensembles is a manifestation of the finite system size, and we investigated this issue further by performing $NVT$ simulations using the NN potential at different density and temperature using a larger system size of 1024 atoms, comparing the results to the aforementioned $NVT$ simulations using 128 atoms in Sec IV B. For each thermodynamic condition, we used both the atomic hydrogen and molecular hydrogen as the starting configurations, and the simulation time of each run is about 100 ps.

As can be seen from Figure S11, at larger system size the molecular fraction under all the conditions considered does not show a drift from the ones obtained at the small system size. However, at low temperatures (600K and 800K), the large 1024 atom system shows a bifurcation behavior close to the transition densities, which is a hysteresis behavior that is typical of first-order transitions. At higher temperatures $T \geq 1000\ K$, the molecular to atomic transition remains smooth. These suggest that at 600K and 800K the hydrogen system experiences a genuine first-order phase transition, and that at higher temperatures the smooth transition suggested by Figure S11 is not due to the finite size effects but is intrinsic to the system.

Another key advantage of employing a large system size is that it is much easier to identify and characterize solid structures in case if the hydrogen system freezes. For the 1024 atom system, as shown in Fig. S12, the atomic configuration at the end of the $NVT$ simulation at 800 K clearly shows evenly spaced planes, and resembles closely-packed and orientationally-disordered molecular hydrogen crystals.

15

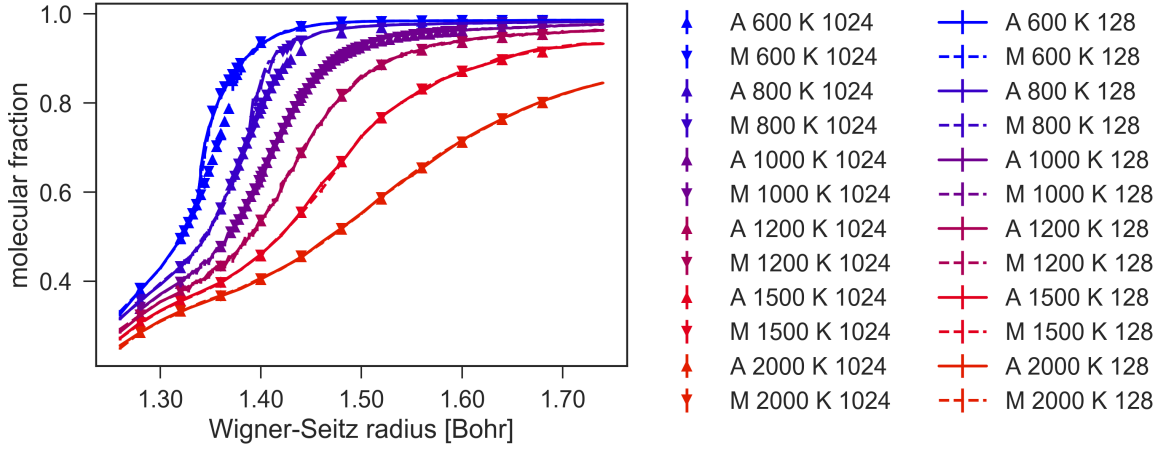

FIG. S11: The comparison of the molecular fraction for the large (1024 atoms) and the small system (128 atoms) at different temperatures and densities.

## B. Finite size effects on simulated quenches

In the AIMD simulations described above in Sec II A, as well in other previous DFT and QMC simulations [5–7], the simulation setup is an $NVT$ simulation using a relatively small system. To further investigate how such setups affect the phase transition behavior, we performed cooling and heating simulations using the NN potential at constant volume using a system size of 128 hydrogen atoms. The Nose-Hoover thermostat were used to control the temperature. The time step was set equal to 0.0002 ps, and the total simulation time in each cooling or heating run was set equal to 400 ps. During the cooling run, the initial temperature is 2000K and final temperature is 0.1K. During the heating run, the initial and the finite temperatures are 100K and 2100K, respectively. The initial configuration of the heating run is the final solid configuration of the cooling simulation at the same constant volume. The evolution of potential energy, pressure and molecular fraction as a function of temperature during the cooling and the heating is plotted in Fig. S13. Similar with the atomic configurations shown in Fig. S12, we observed crystal-like structures after the atomic to molecular transition during the cooling runs. Fig. S13 does not show any hysteresis across the molecular to atomic transition upon changing the temperature, meaning that the transition is almost barrier-less.

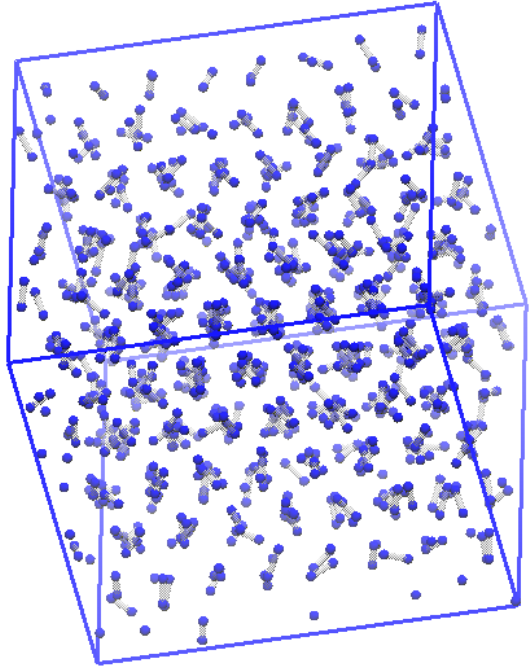

FIG. S12: A snapshot from a MD *NVT* simulation using the MLP, at 800 K at a density corresponding to a Wigner-Seitz radius of 1.44 Bohr.

## VI.  DETAILS OF THE SIMULATIONS DESCRIBED IN THE MAIN TEXT

### A.  Cooling and heating simulations

We perform cooling and heating simulations at constant pressure to estimate an upper and lower bound of the melting temperature. The Nose-Hoover thermostat and isotropic barostat were used to control the temperature and pressure. The time step was set equal to 0.0002 ps, and the total simulation time in each run was set equal to 800 ps. The system size is 1728 atoms. Note that the use of large system size and the number of atoms that fit multiples of 12 are important for crystal structure predictions [13]. During the cooling run, the initial temperature is 1600K and final temperature is 0.1K, and the cooling rate is about 2 K/ps. During the heating run, the initial and the finite temperatures are 100K and 1700K, respectively. The initial configuration of the heating run is the final solid configuration of

17

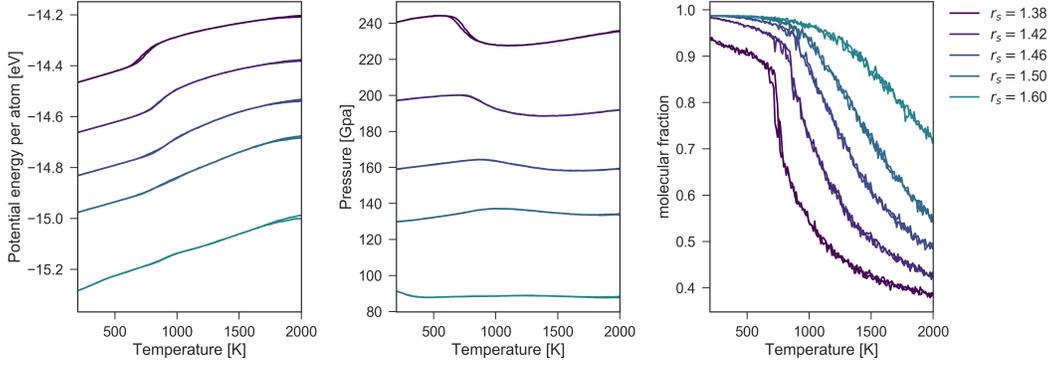

FIG. S13: The evolution of potential energy (left panel), pressure (middle panel), and molecular fraction (right panel) during the cooling and the heating simulations of the 128 H systems under the constant volume condition.

the cooling simulation at the same constant pressure. The evolution of potential energy and molar volume as a function of temperature during the cooling and the heating is plotted in Fig. S14, and the upper and the lower temperatures of the hysteresis are used as a proxy for the upper and the lower bound of the melting point. All the solid configurations obtained from the cooling simulations are provides as a part of the Supplemental Information.

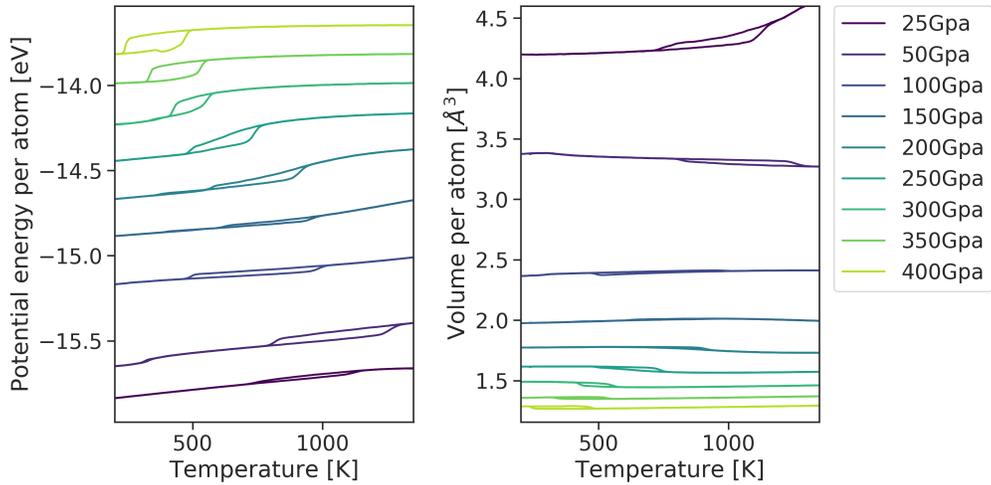

FIG. S14: The evolution of potential energy (left panel) and molar volume (right panel) during the cooling and the heating simulations of the 1728 H systems.

## B. *NPT* simulations

The *NPT* ensemble was employed throughout with the Nose-Hoover thermostat and isotropic barostat used to control the temperature and pressure, as implemented in LAMMPS The time step was set equal to 0.2 fs, which gives a negligible drift in the conserved quantities for this system. Each simulation was run for a total of 400 ps. A supercell with 512 atoms was used, which size is sufficient to give converged results on potential energy, volume and molecular fraction. *NPT* Simulations over a broad range of temperatures ( 250 K-6000K) and pressures (25 GPa -400 GPa ) were performed, with a fine grid of 100 K and 25 GPa , and we used a even smaller grid near the solid-liquid transition and the molecular-atomic transition zones.

We have computed the radial distribution functions for H-H pairs, and in Fig. S15 we show an example at 1500K, which shows a smooth change as pressure increases.

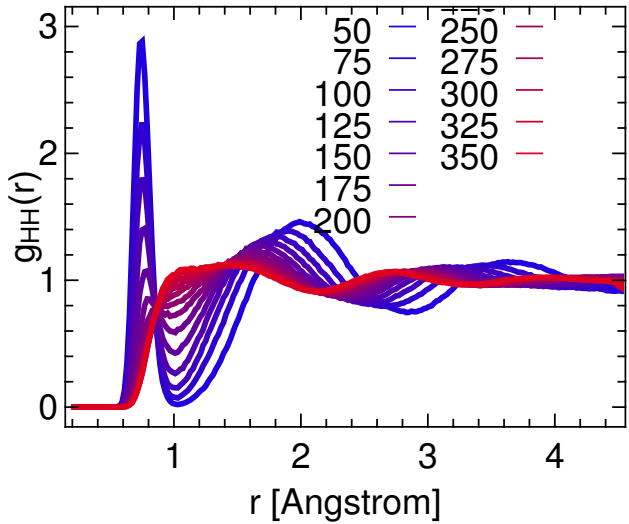

FIG. S15: Radial distribution function at 1500K, computed from *NPT* simulations of 512 H atoms.

## C. Metadynamics simulations

*NPT* Simulations at $T$ =500, 600, 800, 1000, 1200, 1500, 2000, 2500, 3000, 3500, 4000, 4500K and pressures 50GPa -400GPa were performed. The details on the *NPT* simulations are identical to those reported in Section VI B. Each simulation was run for 400 ps. On top

19

of that, well-tempered metadynamics is used to induce transitions between the molecular and the atomic hydrogen system. We used the number of bonded hydrogen atoms, defined as atoms with one neighbor within a smooth cutoff that starts from 0.8 Å and decays to zero at 1.1 Å, as the collective variable. A Gaussian in the collective variable space is deposited every 400 time steps, whose height is gradually diminishing following the well-tempered scheme with the bias-factor set to 200.

The relevant section of a PLUMED [14] is given below

```
COORDINATIONNUMBER ...
LABEL=cn
SPECIES=1-512
SWITCH={CUBIC D_0=0.8 D_MAX=1.1}
MEAN
MORE_THAN1={SMAP R_0=0.5 D_0=0.5 A=8 B=8}
BETWEEN1={GAUSSIAN UPPER=1.2 LOWER=0.8 SMEAR=0.2}
LOWMEM
... COORDINATIONNUMBER


DUMPMULTICOLVAR STRIDE=1000 DATA=cn FILE=cn.xyz

METAD ...
LABEL=metad
ARG=cn.between-1
PACE=400 HEIGHT=2.0 SIGMA=2.0 FILE=HILLS
TEMP=TEMPERATURE BIASFACTOR=200
... METAD
```

### D. Computing electron density of states

We selected 10 de-correlated configurations of H-128 systems from $NVT$ simulations using the MLP, for each density. For each configuration, we first performed a self-consistent

calculations using 4x4x4 k-points on H-128 system, then we used the generated electron density to do another non-self-consistent calculations using a 12x12x12 k-points grid. The rest of the DFT settings are identical with the ones in Sec I. After the non-self-consistent calculations we used a Gaussian smearing of 0.03 eV when computing the electron density of states. In Fig. S16 we show an example of $f(E)$ at 1500K, which shows a smooth change as pressure increases. In Fig. S17 we show the electron occupancy at the Fermi level $f(E_f)$ for each condition.

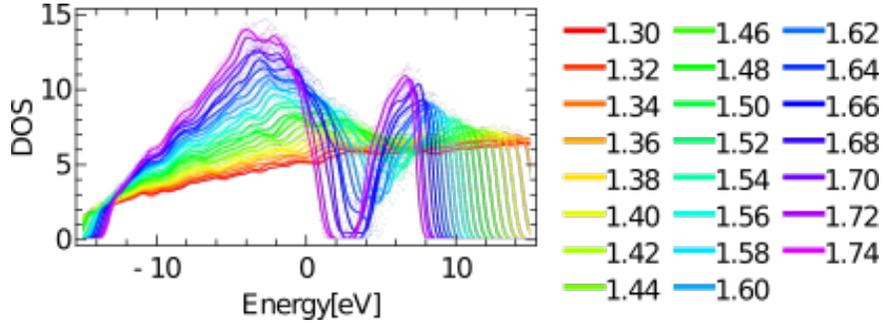

FIG. S16: Electron density of states $f(E)$ at 1500K for 128 atom hydrogen systems at different pressures. Fine dots correspond to individual simulations, while full lines correspond to averages over 10 snapshots at each density described by the Wigner-Seitz radius in Bohr.

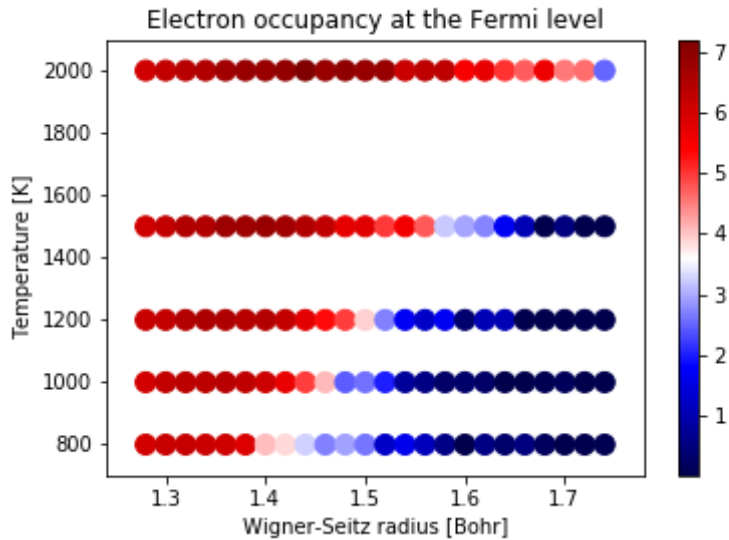

FIG. S17: Electron occupancy at the Fermi level.

## VII. DATA FILES

The machine learning potential for high pressure hydrogen based on PBE DFT, the training set for the potential, and all necessary simulation input files are included in Supplemental Information.


}

\end{document}